\documentclass[aps,prb,twocolumn,amssymb,amsmath,superscriptaddress]{revtex4-1}
\usepackage{bm}
\usepackage{times}
\usepackage{graphicx}
\usepackage{color}
\usepackage{dcolumn}
\usepackage[colorlinks=true, letterpaper=true, pdfstartview=FitV, linkcolor=blue, citecolor=blue, urlcolor=blue]{hyperref}
\usepackage{appendix}
\usepackage[normalem]{ulem}

\begin{document}
\title{Quantum third-order nonlinear Hall effect of a four-terminal device with time-reversal symmetry}
\author{Miaomiao Wei}
\affiliation{College of Physics and Optoelectronic Engineering, Shenzhen University, Shenzhen 518060, China}
\author{Longjun Xiang}
\affiliation{College of Physics and Optoelectronic Engineering, Shenzhen University, Shenzhen 518060, China}
\author{Luyang Wang}
\affiliation{College of Physics and Optoelectronic Engineering, Shenzhen University, Shenzhen 518060, China}
\author{Fuming Xu}
\email[]{xufuming@szu.edu.cn}
\affiliation{College of Physics and Optoelectronic Engineering, Shenzhen University, Shenzhen 518060, China}
\author{Jian Wang}
\email[]{jianwang@hku.hk}
\affiliation{College of Physics and Optoelectronic Engineering, Shenzhen University, Shenzhen 518060, China}
\affiliation{Department of Physics, The University of Hong Kong, Pokfulam Road, Hong Kong, China}

\begin{abstract}
The third-order nonlinear Hall effect induced by Berry-connection polarizability tensor has been observed in Weyl semimetals T$_d$-MoTe$_2$ as well as T$_d$-TaIrTe$_4$. The experiments were performed on bulk samples, and the results were interpreted with the semiclassical Boltzmann approach. Beyond the bulk limit, we develop a quantum nonlinear transport theory to investigate the third-order Hall response of a four-terminal setup with time-reversal symmetry in quantum regime. The quantum nonlinear theory is verified on a model system of monolayer MoTe$_2$, and numerical results on the angle-resolved Hall currents are qualitatively consistent with the experiment. More importantly, quantum signatures of the third-order Hall effect are revealed, which are independent of the system symmetry. The first quantum signature is quantum enhancement of the third-order Hall current, which is characterized by sharp current peaks whose magnitudes are three orders larger than the first-order Hall current. Such quantum enhancement originates from quantum interference in coherent transport, and it can be easily destroyed by dephasing effect. The second quantum signature is disorder-induced enhancement of the third-order Hall current for weak disorders. Our findings reveal quantum characteristics of the third-order Hall effect, and we propose feasible ways to enhance it in nanoscale systems. The quantum third-order theory developed in this work provides a general formalism for describing nonlinear coherent transport properties in multi-terminal devices, regardless of the system symmetry.
\end{abstract}
\maketitle

\section{Introduction}
Berry curvature related nonlinear Hall responses in time-reversal invariant systems have attracted intensive research interest in condensed matter physics as well as material science. The second-order Hall effect induced by Berry curvature dipole (BCD) has been thoroughly studied both theoretically\cite{L-Fu1,Guinea1,NHEREV,Lee,Ortix,Sodemann,Tsymbal} and experimentally in various aspects\cite{S-Xu,Q-Ma,K-Kang,J-Xiao,H-Yang}. Meanwhile, the third-order Hall effect induced by Berry-connection polarizability tensor (BPT), another band geometric quantity\cite{Gao14,Gao15,Gao17,syyang21}, was experimentally realized in thick micron-sized T$_d$-MoTe$_2$ samples in 2021\cite{THE2021}, where the third-order Hall current depends on the relaxation time and hence it is an extrinsic Hall effect. Very recently, room temperature third-order Hall effect was observed in bulk Weyl semimetal TaIrTe$_4$ with thickness of 160 $nm$, which can stably exist for several months\cite{THE2022NSR}. On the theoretical aspect, BPT-induced third-order Hall effect was predicted in multi-Weyl semimetals through model calculation\cite{NLHE2021AN} and monolayer FeSe with first-principles calculation\cite{THE2022}. If the time-reversal symmetry is broken, it was proposed that the third-order nonlinear anomalous Hall effect can be caused by Berry curvature quadruples in certain antiferromagnets such as monolayer SrMnBi$_2$\cite{KTLAW2021}.

Up to now, the third-order Hall effect with time-reversal symmetry was experimentally performed on bulk samples\cite{THE2021,THE2022NSR} to observe optimized nonlinear response signals. Related theoretical interpretations were based on the second-order semiclassical approach\cite{Gao14} for bulk systems, where the field-induced Berry connection and BPT play crucial roles on the extrinsic third-order Hall effect\cite{THE2021,NLHE2021AN,THE2022NSR}. Quantum characteristics of the third-order Hall effect in nanoscale systems has not been addressed. In nanoscale devices, quantum coherent transport is dominant. Therefore, a real-space quantum nonlinear theory is necessary to simulate the third-order Hall response of multi-terminal systems in quantum regime and compare with the second-order semiclassical approach, which is absent so far.

To fill in this gap, we develop a quantum nonlinear transport theory to numerically investigate the third-order Hall response of multi-terminal devices in the presence of time-reversal symmetry. This gauge-invariant quantum nonlinear theory allows direct calculation of the third-order conductances, rather than the semiclassical conductivity. We apply this theory to a 2-dimensional (2D) massive Dirac model describing monolayer T$_d$-MoTe$_2$, and calculate angular-dependent Hall currents (from first order to third order) of a four-terminal Hall setup via rotating four perpendicular leads. The numerical results are qualitatively consistent with both the experiment and the semiclassical approach in Ref.[{\onlinecite{THE2021}}]. More importantly, in quantum regime, the third-order Hall current can be enhanced to three orders of magnitude larger than the first-order Hall current, which was only one order of magnitude in the bulk system\cite{THE2021}. Experimentally observed third-order Hall signals in bulk systems are relatively weak, where the typical third-order Hall voltage is in the order of $\mu V$ in response to a driving voltage in the order of 10 $mV$\cite{THE2021,THE2022NSR}. Therefore, enhanced third-order Hall signal is expected in quantum regime. Such quantum enhancement of the third-order Hall current is attributed to the quantum interference in coherent transport, which is verified by introducing phase relaxation process that severely reduces the sharp current peaks. We also observe disorder-induced enhancement of the third-order Hall effect, where similar disorder effect has been predicted in BCD-induced second-order Hall effect\cite{disorderNHE,QTNHE}. These findings highlight quantum characteristics of the third-order Hall effect, which can be experimentally studied in nanoscale Weyl semimetals T$_d$-MoTe$_2$ and T$_d$-TaIrTe$_4$, etc.

The rest of the paper is organized as follows. In Sec. II, a gauge-invariant quantum third-order transport theory is derived. The quantum nonlinear theory is applied to a four-terminal device with time-reversal symmetry, and numerical results are presented in Sec. III. A summary is finally given in Sec. IV.

\section{Quantum third-order nonlinear transport theory}
We start from the Landauer-B\"{u}ttiker formula describing quantum transport ($\hbar= e=1$)
\begin{eqnarray}
I_\alpha = \sum_\beta \int_E  {\rm Tr}[A_{\alpha\beta}] f_\beta \label{i1},
\end{eqnarray}
where $A_{\alpha\beta}=\Gamma_\alpha G^r \Gamma_{\alpha\beta} G^a$, $\Gamma_{\alpha\beta}=\Gamma_\beta - \delta_{\alpha\beta} \Gamma$, and $\Gamma = \sum_\alpha \Gamma_\alpha$. $G^r(G^a)$ is the retarded(advanced) Green's function. $\Gamma_\alpha$ is the linewidth function of probe $\alpha$ characterizing the coupling between this probe and the scattering region. $f_\beta = f_\beta(E + v_\beta)$ is the Fermi distribution function of probe $\beta$, and $v_\beta$ is the bias voltage in this probe.

One of the fundamental requirements for quantum transport is the gauge-invariant condition\cite{but01,but02}, i.e., the current remains unchanged when the bias voltage in each probe changes by a constant amount. In quantum transport, one has to consider the self-consistent Coulomb interaction to satisfy gauge invariance. This Coulomb potential is due to the charge injection and hence it is a nonequilibrium potential which vanishes at zero bias\cite{but01,but02}. We use a two-probe system to illustrate the importance of gauge invariance. From Eq.(\ref{i1}), the current $I_L$ from the left probe to the second order in bias voltage is expressed as
\begin{eqnarray}
I_L = G_{11} (v_L - v_R) +G_{111} (v_L^2 - v_R^2) + ...
\end{eqnarray}
For $v_L=v$ and $v_R=0$, the current is $I_L = G_{11} v +G_{111} v^2$; while for $v_L = 0$ and $v_R = -v$, the current is $I_L = G_{11} v - G_{111} v^2$, which is clearly gauge dependent and the result is qualitatively incorrect. Therefore, it is essential to have a gauge-invariant formalism in order to obtain reliable results. The gauge-invariant theory of the second-order conductance has been developed long ago\cite{but1,ma,baigen}.
%In addition, in terms of the second-order conductance $G_{\alpha \beta \gamma}$, quantum second-order Hall transport properties of a four-terminal device with time-reversal symmetry was investigated\cite{MWei}, where nonlinear Hall responses similar to the semiclassical BCD-induced second-order Hall effect were revealed. It was found that internal Coulomb interaction and phase relaxation process play important roles in the second-order Hall effect\cite{MWei}.
In this work, we develop the quantum third-order transport theory.

To investigate the nonlinear conductance, it is convenient to work within the nonequilibrium Green's function formalism (NEGF). In this formalism, the Hamiltonian of the scattering region contains a self-consistent Coulomb potential and hence the retarded Green's function is written as
\begin{eqnarray}
G^r(E) = \frac{1}{E- H -q U - \Sigma^r},
\end{eqnarray}
where the self-consistent Coulomb potential satisfies the Poisson equation
\begin{eqnarray}
\nabla^2 U(x) = 4 \pi q i\int_E G^<(E,x,x). \label{poi1}
\end{eqnarray}
The lesser Green's function is $G^< = i G^r \sum_\alpha \Gamma_\alpha f_\alpha G^a$. $\Sigma^r = \sum_{\alpha} \Sigma^r_{\alpha}(E-q v_{\alpha})$ is the self-energy due to the probes\cite{baigen}.

When expanding the current in Eq.(\ref{i1}) to the third order under the condition of small bias voltages, we have
\begin{eqnarray}
I_\alpha &=& G_{\alpha \beta} v_\beta+G_{\alpha\beta\gamma} v_\beta v_\gamma+G_{\alpha \beta \gamma \delta} v_\beta v_\gamma v_\delta+ ... \nonumber \\
         &=& I^{(1)}_\alpha + I^{(2)}_\alpha + I^{(3)}_\alpha + ...,
\end{eqnarray}
where the summation over repeated indices is implied. Here $G_{\alpha \beta}$ ($I^{(1)}_\alpha$), $G_{\alpha \beta \gamma}$ ($I^{(2)}_\alpha$), and $G_{\alpha \beta \gamma \delta}$ ($I^{(3)}_\alpha$) are the linear, second-order, and third-order conductances (currents), respectively. Similarly, when expanding $f_\beta$ in terms of bias voltages and internal potentials, we find
\begin{eqnarray}
f_\beta = f + f' v_\beta +(1/2) f'' v_\beta^2 + (1/6) f''' v_\beta^3 + .... \label{fexp}
\end{eqnarray}
%The third-order nonlinear conductance is defined as
%\begin{eqnarray}
%I_\alpha = G_{\alpha \beta} v_\beta + G_{\alpha \beta \gamma} v_\beta v_\gamma + G_{\alpha \beta \gamma \delta} v_\beta v_\gamma v_\delta,
%\end{eqnarray}
%where the summation over repeated indices is implied. Here $G_{\alpha \beta}$, $G_{\alpha \beta \gamma}$, and $G_{\alpha \beta \gamma \delta}$ are, respectively, the linear, second-order, and third-order conductances.
The gauge-invariant condition requires that
\begin{eqnarray}
&&\sum_\beta G_{\alpha \beta} =0, ~~~ \sum_\beta G_{\alpha \beta \gamma} = \sum_\gamma G_{\alpha \beta \gamma}=0,  \\ \label{eq7}
&&\sum_\beta G_{\alpha \beta \gamma \delta} = \sum_\gamma G_{\alpha \beta \gamma \delta} = \sum_\delta G_{\alpha \beta \gamma \delta} =0. \label{eq8}
\end{eqnarray}
Since $A_{\alpha\beta}$ in Eq.(\ref{i1}) depends on the bias voltage through the Coulomb potential $U$ in $G^r$, we expand $A_{\alpha\beta}$ in terms of the Coulomb potential and find
\begin{eqnarray}
A_{\alpha\beta} = A_{\alpha\beta 0} - \partial_U A_{\alpha\beta} + (1/2) \partial^2_U A_{\alpha\beta} + ..., \label{aexp}
 \nonumber
\end{eqnarray}
where $\partial_U A_{\alpha\beta} \equiv \Gamma_\alpha G^r U G^r \Gamma_{\alpha\beta} G^a + \Gamma_\alpha G^r \Gamma_{\alpha\beta} G^a U G^a$ is the functional derivative\cite{Gasparian}. Now we discuss how to solve $U$ from the Poisson equation. For the third-order conductance $G_{\alpha \beta \gamma \delta}$, it is enough to expand $U$ to the second order in bias
\begin{eqnarray}
U = u_\alpha v_\alpha + u_{\alpha \beta} v_\alpha v_\beta,
\end{eqnarray}
where $u_\alpha$ ($u_{\alpha \beta}$) is the linear (second-order) characteristic potential satisfying the following equations:
\begin{eqnarray}
&& -\nabla^2 u_{\alpha}(x) + 4\pi q^2 \frac{dn(x)}{dE} u_{\alpha}(x) = 4\pi q^2 \frac{dn_\alpha(x)}{dE}, \\
&& -\nabla^2 u_{\alpha \beta} +4\pi q^2 \frac{dn}{dE} u_{\alpha \beta} = 4\pi q^2 \frac{d\tilde{n}_{\alpha \beta}}{dE},
\end{eqnarray}
where
\begin{eqnarray}
\frac{d\tilde{n}_{\alpha \beta}}{dE} = \frac{d^2n_\alpha}{dE^2} \delta_{\alpha \beta} -\frac{d^2n_\alpha}{dE^2} u_\beta -
\frac{d^2n_\beta}{dE^2} u_\alpha +\frac{d^2n}{dE^2} u_\alpha u_\beta. \nonumber
\end{eqnarray}
Here $dn_\alpha/dE$ is the local density of states (DOS) defined as\cite{but1,ywei}
\begin{eqnarray}
\frac{dn_{\alpha}(x)}{dE} = -\int \frac{dE}{2\pi} \partial_E f [G^r_0 \Gamma_{\alpha} G^a_0]_{xx},
\end{eqnarray}
and $dn/dE = \sum_\alpha dn_\alpha/dE$ is the total DOS. The second-order DOS is defined as $d^2n_\alpha/dE^2 = \partial_E dn_\alpha/dE$. Due to the gauge invariance, we have
\begin{eqnarray}
&& \sum_\alpha u_\alpha =1, \nonumber \\
&& \sum_\alpha u_{\alpha \beta} = \sum_\beta u_{\alpha \beta} = 0. \nonumber
\end{eqnarray}
From Eqs.(\ref{i1}), (\ref{fexp}), and (\ref{aexp}), the third-order current is expressed in terms of the third-order conductance as
\begin{eqnarray}
I^{(3)}_\alpha = \sum_\beta \int_E  {\rm Tr}[B_1 +B_2 +B_3].  \nonumber
\end{eqnarray}
Here
\begin{eqnarray}
B_1 =(1/6)  f' v_\beta^3 \partial^2_E A_{\alpha\beta}  = (1/3) f' v_\beta^3 C_{1\alpha \beta},
\end{eqnarray}
where we have dropped the subscript $0$ in $A_{\alpha\beta}$ and performed integration by parts twice with respect to $E$. $C_{1\alpha \beta}$ is expressed as
\begin{eqnarray}
C_{1\alpha \beta} = \Gamma_\alpha (G^r)^2 g_{\alpha\beta} + \Gamma_\alpha g_{\alpha\beta} (G^a)^2 + \Gamma_\alpha G^r g_{\alpha\beta} G^a,  \nonumber
\end{eqnarray}
where $g_{\alpha\beta} = G^r \Gamma_{\alpha\beta} G^a$ and we have used the fact that
\begin{eqnarray}
-\partial_E A_{\alpha \beta} = \Gamma_\alpha G^r G^r \Gamma_{\alpha\beta} G^a + \Gamma_\alpha G^r \Gamma_{\alpha\beta} G^a G^a. \nonumber
\end{eqnarray}
$B_2$ is defined as
\begin{eqnarray}
B_2 &=& (1/2)  f' v_\beta^2 \partial_E \partial_U A_{\alpha\beta}U  - f' v_\beta \partial_U A_{\alpha\beta} U \nonumber \\
&=& -(1/2) f' v_\beta^2 C_{2\alpha \beta\gamma} v_\gamma + f' v_\beta {\tilde C}_{2\alpha \beta\gamma\delta} v_\gamma v_\delta,
\end{eqnarray}
where
\begin{eqnarray}
C_{2\alpha \beta \gamma} &=& \Gamma_\alpha G^r G^r u_\gamma g_{\alpha\beta} + \Gamma_\alpha G^r u_\gamma G^r g_{\alpha\beta}  \nonumber \\
&+& \Gamma_\alpha G^r u_\gamma g_{\alpha\beta} G^a + \Gamma_\alpha G^r g_{\alpha\beta} u_\gamma G^a \nonumber \\
&+& \Gamma_\alpha g_{\alpha\beta} G^a u_\gamma G^a + \Gamma_\alpha g_{\alpha\beta} u_\gamma G^a G^a, \nonumber
\end{eqnarray}
and
\begin{eqnarray}
{\tilde C}_{2\alpha \beta\gamma\delta} &=& \Gamma_\alpha G^r G^r u_{\gamma\delta} g_{\alpha\beta} + \Gamma_\alpha G^r u_{\gamma\delta} G^r g_{\alpha\beta}  \nonumber \\
&+& \Gamma_\alpha G^r u_{\gamma\delta} g_{\alpha\beta} G^a + \Gamma_\alpha G^r g_{\alpha\beta} u_{\gamma\delta} G^a \nonumber \\
&+& \Gamma_\alpha g_{\alpha\beta} G^a u_{\gamma\delta} G^a + \Gamma_\alpha g_{\alpha\beta} u_{\gamma\delta} G^a G^a. \nonumber
\end{eqnarray}
$B_3$ is defined as
\begin{eqnarray}
B_3 =(1/2)  f' v_\beta \partial^2_U A_{\alpha\beta} U^2  =  f' v_\beta \sum_{\gamma \delta} C_{3\alpha \beta \gamma \delta} v_\gamma v_\delta,
\end{eqnarray}
where
\begin{eqnarray}
C_{3\alpha \beta \gamma \delta} &=& \Gamma_\alpha G^r u_\gamma G^r u_\delta g_{\alpha\beta} + \Gamma_\alpha G^r u_\gamma g_{\alpha\beta} u_\delta G^a  \nonumber \\
&+& \Gamma_\alpha g_{\alpha\beta} u_\gamma G^a u_\delta G^a.
\end{eqnarray}
In terms of $C_{i}$, the nonsymmetrized third-order conductance is defined as
\begin{eqnarray}
{\tilde G}_{\alpha \beta \gamma \delta} &=& \int_E f' {\rm Tr}[(1/3) C_{1\alpha\beta}\delta_{\beta \gamma} \delta_{\beta \delta} + {\tilde C}_{2\alpha\beta\gamma\delta}\nonumber \\
&-&(1/2) C_{2\alpha\beta\gamma} \delta_{\beta \delta} + C_{3\alpha\beta\gamma\delta}].\label{eq7}
\end{eqnarray}
It is easy to show the following three relations\cite{fnote1}:
\begin{eqnarray}
&& \sum_\beta \Gamma_{\alpha\beta} = 0, \\
&& \sum_\gamma C_{2\alpha \beta \gamma} = 2 C_{1\alpha \beta }, \\
&& \sum_\delta (C_{3\alpha \beta \delta \gamma} + C_{3\alpha \beta \gamma \delta}) = C_{2\alpha \beta \gamma}.
\end{eqnarray}
%\begin{eqnarray}
%\sum_\gamma C_{2\alpha \beta \gamma} = 2 C_{1\alpha \beta },
%\end{eqnarray}
%and
%\begin{eqnarray}
%\sum_\delta (C_{3\alpha \beta \delta \gamma} + C_{3\alpha \beta \gamma \delta})=  C_{2\alpha \beta \gamma},
%\end{eqnarray}
From these relations, it is straightforward to show that
\begin{eqnarray}
\sum_\beta ({\tilde G}_{\alpha \beta \gamma \delta} + {\tilde G}_{\alpha  \gamma \delta \beta} + {\tilde G}_{\alpha \gamma \beta \delta})
= C_{2\alpha \gamma \delta } - C_{2\alpha \delta \gamma}. \label{eq11}
\end{eqnarray}
Interchanging $\gamma$ with $\delta$ in Eq.(\ref{eq11}) and adding it to Eq.(\ref{eq11}), we finally have
\begin{eqnarray}
\sum_\beta  G_{\alpha \beta \gamma \delta}=0, \label{eq12}
\end{eqnarray}
where $G_{\alpha \beta \gamma \delta}$ is the symmetrized third-order conductance defined as
\begin{eqnarray}
G_{\alpha\beta\gamma\delta} &=& \frac{1}{6}({\tilde G}_{\alpha\beta\gamma\delta} + {\tilde G}_{\alpha\gamma\delta\beta} + {\tilde G}_{\alpha\gamma\beta\delta} \nonumber \\
&+& {\tilde G}_{\alpha \beta \delta \gamma} + {\tilde G}_{\alpha \delta \gamma \beta} + {\tilde G}_{\alpha \delta \beta \gamma}).
\end{eqnarray}
Since $G_{\alpha \beta \gamma \delta}$ is symmetrized, Eq.(\ref{eq12}) leads to Eq.(\ref{eq8}) which is the gauge-invariant condition we wish to obtain. Note that for the third-order conductance, the induced Coulomb potential and characteristic potentials are in the second order of the bias voltage, while for the second-order conductance it is enough to consider the first-order potentials\cite{but1,ma,baigen,ywei}. In this quantum third-order nonlinear transport theory, the characteristic potentials induced by external bias voltages, play similar roles to those of the field-induced Berry connection and BPT in the second-order semiclassical approach.

%In this quantum third-order transport theory, the symmetrized third-order conductance $G_{\alpha \beta \gamma \delta}$ is defined as
%\begin{eqnarray}
%G_{\alpha\beta\gamma\delta} &=& \frac{1}{6}({\tilde G}_{\alpha\beta\gamma\delta} + {\tilde G}_{\alpha\gamma\delta\beta} + {\tilde G}_{\alpha\gamma\beta\delta} \nonumber \\
%&+& {\tilde G}_{\alpha \beta \delta \gamma} + {\tilde G}_{\alpha \delta \gamma \beta} + {\tilde G}_{\alpha \delta \beta \gamma}),
%\end{eqnarray}
%which guarantees gauge invariance. ${\tilde G}_{\alpha\beta\gamma\delta}$ and relevant quantities are listed in the supplemental
% material\cite{supple}.

\begin{figure}[tbp]
\includegraphics[width=\columnwidth]{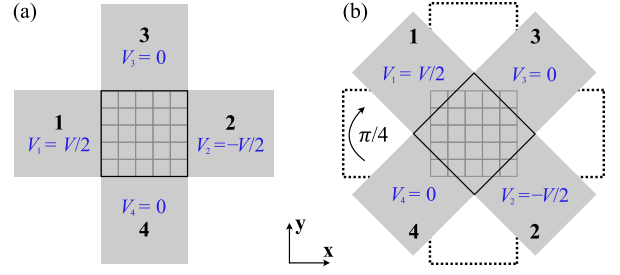}
\caption{ The schematic of a four-terminal Hall setup (a). Its four probes can be rotated by multiples of $\pi/4$, as shown in (b). The symmetry and lattice of the central scattering region (mesh grid) remain the same during the rotation.}\label{fig1}
\end{figure}

For the four-terminal Hall setup shown in Fig.\ref{fig1}, when applying bias voltages on terminals 1 and 2, the quantum third-order Hall current is expressed in terms of the third-order conductances as
\begin{eqnarray}
I^{(3)}_H = I^{(3)}_3 - I^{(3)}_4 = \sum\limits_{\alpha \beta \gamma }[{{{G}_{3\alpha \beta \gamma} - {G}_{4\alpha \beta \gamma}}}]{{V}_{\alpha }}{{V}_{\beta }}{{V}_{\gamma }}.
\end{eqnarray}
In the following, we apply this quantum nonlinear transport theory to study the third-order Hall response of a 2D effective model with time-reversal symmetry (TRS), and compare with experimental results for bulk samples.

\section{Numerical results and discussion}

In the experiment performed on bulk T$_d$-MoTe$_2$\cite{THE2021}, the angular-resolved third-order Hall response was measured via rotating four perpendicular electrodes of a circular device. Due to the $Pmn2_1$ symmetry of the bulk system, the third-order Hall voltage was zero for $\theta =0, \pi/2, \pi$, etc, as shown in Fig.3c of Ref.[{\onlinecite{THE2021}}]. We propose that, one can break the mirror symmetry of MoTe$_2$ so that the third-order Hall response naturally exists for $\theta =0$, i.e., without rotating the electrodes. In the following, we will verify this point with both quantum nonlinear theory and the semiclassical approach.

We study the following 2D massive Dirac Hamiltonian preserving TRS:
\begin{eqnarray}
H\left(\mathbf{k}\right)=A{{k}^{2}}+\left( B{{k}^{2}}+\delta \right){{\sigma }_{z}}+ {{v}_{2}}{{k}_{y}}{{\sigma}_{y}}+ d_0 {{\sigma }_{x}} + \alpha k_x \sigma_y ,\nonumber \label{ham}
\end{eqnarray}
where $A$, $B$, $\delta$, $v_2$, and $d_0$ are system parameters. Here $d_0$ determines the band gap. The presence of $v_2$ breaks the inversion symmetry and leaves single mirror symmetry ${\cal{M}}_x$ in the system. This Hamiltonian captures the key feature of monolayer T$_d$-MoTe$_2$ and T$_d$-WTe$_2$, and it is widely adopted to simulate Berry curvature related physics, including BCD and BPT. The $\alpha$ term effectively breaks the ${\cal{M}}_x$ symmetry. When $\alpha=0$, this system has a BCD aligning along the x-direction due to the ${\cal{M}}_x$ symmetry. Reference [\onlinecite{THE2021}] has demonstrated the BPT tensor distribution of this model in momentum space. In general, if Berry connection is zero, there is no Berry curvature related physics; if Berry curvature is zero with nonzero Berry connection, then BCD is zero and BPT is generally nonzero. Depending on the $\alpha$ term, we have two cases, $\alpha =0$ and $\alpha \neq 0$, which will be discussed in detail below. In the calculation, we conveniently set $A=0$, $B=1$, $\delta=-0.25$, $d_0=0.1$\cite{L-Fu1}, and $\alpha=0.1$.

\begin{figure}
\includegraphics[width=0.85\columnwidth]{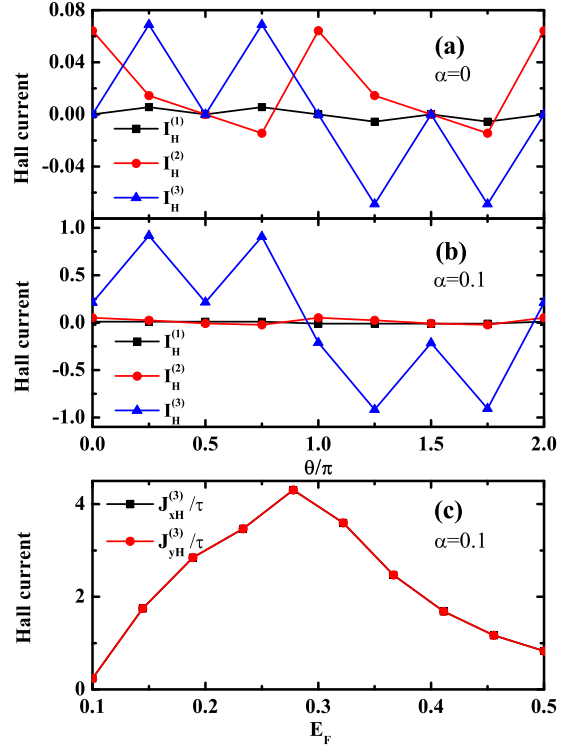}
\caption{ Angle-resolved Hall currents for the four-terminal system with $\alpha=0$ in (a) and $\alpha = 0.1$ in (b). (c) shows the energy dependence of the third-order Hall current from the semiclassical approach\cite{THE2022} without rotating the electrodes, and the current is scaled with the relaxation time $\tau$. }\label{fig2}
\end{figure}

Described by this Hamiltonian, the four-terminal system under investigation is shown in Fig.\ref{fig1}. In the angle-resolved measurement, four perpendicular electrodes were rotated and the Hall current was measured as a function of the rotating angle $\theta$\cite{THE2021}. We follow the same setup in the theoretical calculation by rotating the probes while maintaining the symmetry of the central scattering region (mesh grid), as was done experimentally. Since it is difficult to deal with self-energies of the probes, only rotating angles in the increment of $\pi/4$ are considered. Numerical results on the angle-resolved Hall currents are presented in Fig.\ref{fig2} for the following two cases, where we fix the Fermi energy $E_F=0.15$ and bias voltage difference $V=0.1$.

\noindent{\it Case 1: $\alpha =0$}. In case 1, BCD exists in the system. From Fig.\ref{fig2}(a), we observe the following: (a) The first-order Hall current is suppressed by TRS but it is nonzero due to conductivity anisotropy, and its angle-dependent properties have been theoretically explained in the supplementary of Ref. [\onlinecite{Du2018}] and confirmed by experiments\cite{K-Kang,THE2021}; (b) The BCD-induced second-order Hall current is significant in this 2D system, since mirror symmetry ${\cal{M}}_x$ is preserved; (c) The third-order Hall current $I^{(3)}_H$ is zero for $\theta=0, \pi/2, \pi$ due to the system symmetry, and it is nonzero after $\pi/4$ rotation of the leads. $I^{(3)}_H$ is approximately one order of magnitude larger than the linear order Hall current $I^{(1)}_H$. These results qualitatively agree with the experiment for thick micron-scaled samples\cite{THE2021}. Notice that different orders of Hall responses are distinguishable in experiments through the phase lock-in technique\cite{Q-Ma,K-Kang,H-Yang,THE2021}.

\noindent{\it Case 2: $\alpha \neq 0$}. In case 2, the $\alpha$ term breaks the ${\cal{M}}_x$ symmetry, and the system is a general noncentrosymmetric one with TRS. As one can expect, Fig.\ref{fig2}(b) shows that the second-order Hall current almost vanishes due to the breaking of ${\cal{M}}_x$. BPT is less influenced, and the third-order Hall current is the dominant response in this case. Specifically, $I^{(3)}_H$ is nonzero at $\theta=0, \pi/2, \pi$, which means that the third-order Hall signal is observable in this 2D system without rotation of leads.

To further verify this point, we calculate the third-order Hall response $J^{(3)}_H$ with the semiclassical approach derived in Ref.[\onlinecite{THE2022}] and applied to this particular Hamiltonian [Eq.(\ref{ham})]. Prediction from this semiclassical approach is presented in the Appendix. Numerical results illustrated in Fig.\ref{fig2}(c) clearly show that the third-order Hall effect can exist in this 2D system without rotating the electrodes when the mirror symmetry is broken. It is found that the third-order Hall signals are prominent in a large energy range. As argued in Ref.[\onlinecite{THE2022}], in the calculation we also consider only the third-order response in linear order of the relaxation time $\tau$, since it is induced by BPT and can be more precisely extracted from experimental signals. We point out that the semiclassical approach deals with the conductivity of bulk systems, while the quantum nonlinear transport theory considers the conductances among different probes of the four-terminal setup. The quantum nonlinear transport theory also allows us to investigate the third-order Hall effect in quantum regime, which is beyond the reach of the semiclassical approach.

\begin{figure}[tbp]
\includegraphics[width=\columnwidth]{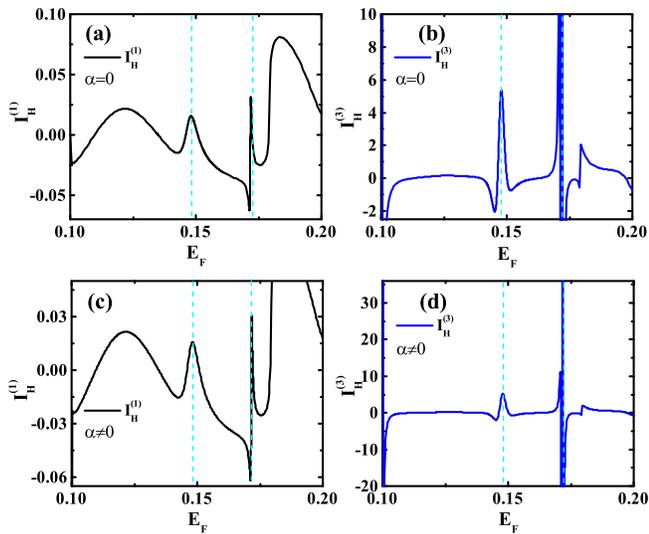}
\caption{The linear and third-order Hall currents as a function of the Fermi energy. The rotating angle is $\pi/4$ for all panels. $\alpha=0$ in (a) and (b), and $\alpha = 0.03$ in (c) and (d).}\label{fig3}
\end{figure}

In the following, we study quantum characteristic of the third-order Hall current in coherent transport. In Fig.\ref{fig3}, we plot the linear and third-order Hall currents versus the Fermi energy at the rotating angle $\theta=\pi/4$. It is found that both $I^{(1)}_H$ and $I^{(3)}_H$ changes with the Fermi energy in large ranges, and sharp current peaks are observed. As shown by the dashed lines, there is precise correspondence between the sharp peaks of $I^{(1)}_H$ and those of $I^{(3)}_H$. Remarkably, $I^{(3)}_H$ is significantly enhanced around the narrow peaks of $I^{(1)}_H$. The narrower the $I^{(1)}_H$ peak, the larger the $I^{(3)}_H$. The maximum value of $I^{(3)}_H$ is over 1300 in Fig.\ref{fig3}(b), which is three to four orders of magnitude larger than $I^{(1)}_H$. These current peaks are similar to the resonant transmission peaks in two-probe systems, which is due to the quantum interference in coherent transport. The quantum enhancement of $I^{(3)}_H$ is the first quantum signature of the third-order Hall response in nanoscale systems. This common feature exists regardless of the presence or absence of mirror symmetry, suggesting that it is independent of the system symmetry.

\begin{figure}
\includegraphics[width=0.9\columnwidth]{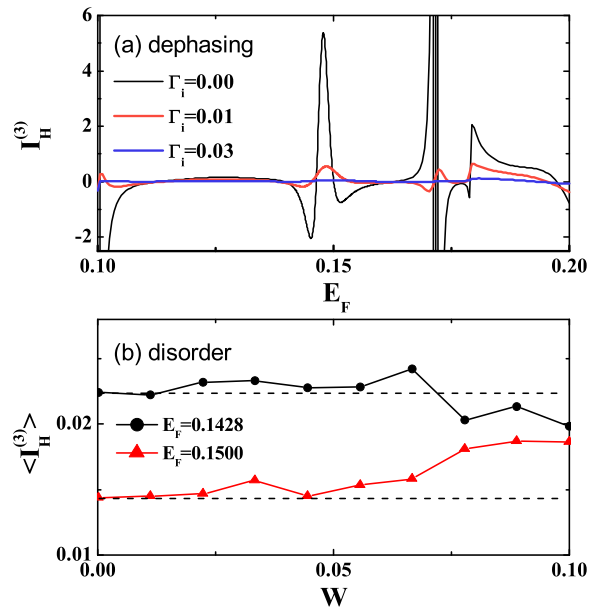}
\caption{(a) The effect of dephasing on the third-order Hall current for several dephasing strengths. (b) Disorder-induced enhancement of the third-order Hall current for certain energies. }\label{fig4}
\end{figure}

It is well known that, coherent transport can be destroyed by the dephasing effect or the phase relaxation process\cite{but2,wang93}, such as thermal broadening, where sharp resonant peaks due to quantum interference are suppressed or smeared out. To further demonstrate the quantum nature of the $I^{(3)}_H$ peaks, we introduce dephasing mechanism into the system via the virtual probe technique\cite{but2,but3,datta}. For simplicity, we assume that the dephasing effect only exists in the central scattering region and a virtual voltage probe is attached to each site with the zero-current constraints. The retarded self-energy of the virtual probe is $\Sigma_i^r = -i \Gamma_i/2$, with $\Gamma_i$ being the dephasing strength. After solving the nonlinear current-voltage equations for all the real and virtual probes, we obtain the third-order Hall current of the four-terminal system in the presence of dephasing, and numerical results are shown in Fig.\ref{fig4}(a). We see that the sharp current peaks due to quantum interference are greatly reduced for the dephasing strength $\Gamma_i=0.01$, and coherent transport is completely suppressed for $\Gamma_i=0.03$. This numerical evidence proves the quantum nature of the sharp $I^{(3)}_H$ peaks.

We also investigate the effect of disorder on the third-order Hall current. Anderson-type disorder with strength $W$ is added into the central region as random on-site energies, and the ensemble-averaged current for several Fermi energies are displayed in Fig.\ref{fig4}(b). We find that, for certain energies, $\langle I^{(3)}_H \rangle$ increases with the disorder strength, which corresponds to disorder enhancement of the quantum third-order Hall current. Such enhancement is also not affected by the mirror symmetry. Similar disorder enhancement has been reported on the second-order Hall effect induced by BCD, with both semiclassical\cite{disorderNHE} and quantum approaches\cite{QTNHE}. The quantum diagrammatic theory predicted that\cite{QTNHE}, disorder effect plays decisive role in quantum second-order Hall transport. Here we numerically demonstrate that there is also disorder enhancement of the third-order Hall effect induced by BPT, which is another quantum signature of the third-order nonlinear Hall effect in quantum regime. Since both quantum interference and disorder effect are common features in quantum transport and not affected by system symmetry, it is expected that these enhancements of the third-order Hall effect widely exist in nanoscale systems.

We have studied the BCD-induced second-order Hall effect in terms of the second-order conductance\cite{MWei}, and found one-to-one corresponding nonlinear transport properties similar to the semiclassical prediction\cite{L-Fu1}. In this work, we develop the quantum third-order nonlinear transport theory, and apply it to investigate the BPT-induced third-order Hall effect\cite{THE2021}. The numerical results qualitatively agree with experimental observations performed on bulk systems in micron scale\cite{THE2021}. Quantum signatures of the third-order Hall effect are revealed, which are characterized by quantum enhancement induce by quantum interference and disorder effect. The success of this quantum nonlinear transport theory in describing the second-order and third-order nonlinear Hall effects shows that it is suitable for describing nonlinear transport properties, especially in quantum regime. We emphasize that, this quantum nonlinear transport theory is a general formalism for investigating nonlinear transport through multi-terminal systems, regardless of the system symmetry.

\section{Conclusion}
In summary, we have developed a quantum nonlinear transport theory to study the third-order nonlinear Hall effect of a 2D four-terminal system with time-reversal symmetry. Angle-resolved Hall currents obtained from the quantum nonlinear theory on a model monolayer MoTe$_2$ system are qualitatively consistent with the experimental results for bulk MoTe$_2$. More importantly, it is found that in coherent transport, the third-order Hall current can be significantly enhanced by quantum interference to three orders of magnitude larger than the first-order Hall signal. Such quantum enhancement is vulnerable to dephasing effect. We also find disorder-induced enhancement of the third-order Hall current in quantum transport. These numerical findings highlight quantum characteristics of the third-order Hall effect induced by BPT, and we expect that experimental observations can be carried out on platforms such as nanoscale MoTe$_2$ and TaIrTe$_4$.

\section*{acknowledgments}
We acknowledge support from the National Natural Science Foundation of China (Grants No. 12034014, No. 12174262, and No. 12004442), Natural Science Foundation of Guangdong (Grant No. 2020A1515011418) and Natural Science Foundation of Shenzhen (Grant No. 20200812092737002).

\section{Appendix: The semiclassical approach for the third-order current}
According to the extended semiclassical approach in Ref.[\onlinecite{THE2022}], the third-order current is expressed as ($\hbar = e =1$)
\begin{equation}
\begin{split}
{{J}^{\left( 3 \right)}}=&-\tau \int{\left[ d\textbf{k} \right]}{{\nabla }_{\textbf{k}}}\varepsilon \left( \textbf{E}\cdot {{\nabla }_{\textbf{k}}} \right)\left[ {{\varepsilon }^{\left( 2 \right)}}f_{0}^{'} \right] \nonumber \\
& -\tau \int{\left[ d\textbf{k} \right]}{{\nabla }_{\textbf{k}}}{{\varepsilon }^{\left( 2 \right)}}\left( \textbf{E}\cdot {{\nabla }_{\textbf{k}}} \right){{f}_{0}} \nonumber \\
& -\tau \int{\left[ d\textbf{k} \right]}\textbf{E}\times {{\mathbf{\Omega} }^{\left( 1 \right)}}\left( \textbf{E}\cdot {{\nabla }_{\textbf{k}}} \right){{f}_{0}} \nonumber \\
& - {\tau ^3}\int {\left[ {d\textbf{k}} \right]} {\nabla _\textbf{k}}\varepsilon {\left( {\textbf{E} \cdot {\nabla _\textbf{k}}} \right)^3}{f_0},
\end{split}
\end{equation}
where ${{\varepsilon }^{\left( 2 \right)}}=-\frac{1}{2}\textbf{E}\overset{\lower0.5em\hbox{$\smash{\scriptscriptstyle\leftrightarrow}$}} {G}\textbf{E}$ is the second-order correction to energy, ${{G}_{ab}}=2\operatorname{Re}\sum\limits_{n\ne 0}{\frac{{{\left( {{A}_{a}} \right)}_{0n}}{{\left( {{A}_{b}} \right)}_{n0}}}{{{\varepsilon }_{0}}-{{\varepsilon }_{n}}}}$ is the Berry-connection polarizability tensor (BPT) with the interband Berry connection ${{\left( {{A}_{a}} \right)}_{mn}}=\left\langle {{u}_{m}}\left| i{{\partial }_{k_a}}\left| {{u}_{n}} \right. \right. \right\rangle $, and ${{\Omega }^{\left( 1 \right)}}={{\nabla }_{\textbf{k}}}\times \overset{\lower0.5em\hbox{$\smash{\scriptscriptstyle\leftrightarrow}$}} {G}\textbf{E}$ is the field-induced Berry curvature.

Considering only the linear term in $\tau$ since the third-order term in $\tau$ is a Drude-like contribution, the third-order current is simplified as
\begin{equation}
\begin{split}
  {{J}^{\left( 3 \right)}}=&-\tau [\int{\left[ d\mathbf{k} \right]}\frac{\hbar }{2}{{v}_{0}}\left( \mathbf{E}G\mathbf{E} \right)\left( \mathbf{E}\cdot {{\partial }_{\mathbf{k}}} \right){{f}_{0}}^{\prime } \\
 & +\int{\left[ d\mathbf{k} \right]}{{\nabla }_{\mathbf{k}}}\times {{\mathbf{\Omega }}^{\left( 1 \right)}}\left( \mathbf{E}\cdot {{\partial }_{\mathbf{k}}} \right){{f}_{0}}]. \\
\end{split}
\end{equation}

For the 2D Dirac model defined in Eq.(\ref{ham}), the current along $y$ direction is
\begin{equation}
\begin{split}
  J_{y}^{\left( 3 \right)}= &-\tau [\int{\left[ d\mathbf{k} \right]}\frac{\hbar }{2}{{v}_{0}}\left( \mathbf{E}G\mathbf{E} \right)\left( \mathbf{E}\cdot {{\partial }_{\mathbf{k}}} \right){{f}_{0}}^{\prime } \\
 & +\int{\left[ d\mathbf{k} \right]}{{\left( {{\nabla }_{\mathbf{k}}}\times {{\mathbf{\Omega }}^{\left( 1 \right)}} \right)}_{y}}\left( \mathbf{E}\cdot {{\partial }_{\mathbf{k}}} \right){{f}_{0}}] \\
  =&-\tau [\int{\left[ d\mathbf{k} \right]}{{\chi }_{yxxx}}E_{x}^{3}+{{\chi}_{yxxy}}E_{x}^{2}{{E}_{y}}\\
  &+{{\chi }_{yxyy}}{{E}_{x}}E_{y}^{2} +{{\chi }_{yyyy}}E_{y}^{3}] \\
\end{split}
\end{equation}
where
\begin{equation}
\begin{split}
{\chi _{yxxx}} =& \left( {{\partial _y}{G_{xx}} - {\partial _x}{G_{xy}}} \right){\partial _x}{f_0} + \frac{\hbar }{2}{v_0}{G_{xx}}{\partial _x}{f_0}', \nonumber \\
{\chi _{yxxy}} =& \left( {{\partial _y}{G_{xy}} - {\partial _x}{G_{yy}}} \right){\partial _x}{f_0} + \left( {{\partial _y}{G_{xx}} - {\partial _x}{G_{xy}}} \right){\partial _y}{f_0} \\
+& \frac{\hbar }{2}{v_0}\left( {2{G_{xy}}{\partial _x}{f_0}' + {G_{xx}}{\partial _y}{f_0}'} \right),\nonumber \\
{\chi _{yxyy}} =& \left( {{\partial _y}{G_{xy}} - {\partial _x}{G_{yy}}} \right){\partial _y}{f_0} \\
+& \frac{\hbar }{2}{v_0}\left( {{G_{yy}}{\partial _x}{f_0}' + 2{G_{xy}}{\partial _y}{f_0}'} \right), \nonumber \\
{\chi _{yyyy}} =& \frac{\hbar }{2}{v_0}{G_{yy}}{\partial _y}{f_0}'.
\end{split}
\end{equation}

When an electric field is applied in $x$ direction, the third-order Hall current along $y$ direction is
\begin{equation}
J_{yH}^{\left( 3 \right)}=-\tau \int{\left[ d\textbf{k} \right]}{{\chi }_{yxxx}}E_{x}^{3}.
\end{equation}

Similarly, for an electric field along $y$ direction, the Hall current along $x$ direction is
\begin{equation}
J_{xH}^{\left( 3 \right)}=-\tau \int{\left[ d\textbf{k} \right]}{{\chi }_{xyyy}}E_{y}^{3},
\end{equation}
where ${{\chi }_{xyyy}}=\left( {{\partial }_{x}}{{G}_{yy}}-{{\partial }_{y}}{{G}_{xy}} \right){{\partial }_{x}}{{f}_{0}}+\frac{\hbar }{2}{{v}_{0}}{{G}_{yy}}{{\partial }_{y}}{{f}_{0}}'$. In the calculation, we set $E_x = E_y =1$ and scale the current with $\tau$. Then the third-order Hall currents are further simplified as
\begin{align}
J^{(3)}_{xH} / \tau &= -\int{\left[ d\textbf{k} \right]}{{\chi }_{xyyy}}, \nonumber \\
J^{(3)}_{yH} / \tau &= -\int{\left[ d\textbf{k} \right]}{{\chi }_{yxxx}}.
\end{align}
Apparently, the third-order Hall current naturally exists in this noncentrosymmetric system.

\end{document}